\documentclass[final,5p,times,twocolumn,numbers,sort&compress]{elsarticle}

\usepackage{graphicx}
\usepackage{amsmath}
\usepackage{subfigure}
\usepackage{wrapfig}
\usepackage{epsfig}
\usepackage{float}
\usepackage{color}
\usepackage[section]{placeins}

\usepackage[utf8]{inputenc}
\usepackage{booktabs}
\usepackage{xcolor}
\definecolor{darkred}{rgb}{0.4,0.0,0.0}
\definecolor{darkgreen}{rgb}{0.0,0.4,0.0}
\definecolor{darkblue}{rgb}{0.0,0.0,0.4}
\usepackage[bookmarks,linktocpage,colorlinks,
    linkcolor = darkred,
    urlcolor  = darkblue,
    citecolor = darkgreen]{hyperref}
    
\usepackage{comment}
\newcommand{\be}{\begin{equation}}
\newcommand{\ee}{\end{equation}}

\usepackage{caption} 
\def\msbar{\overline{\rm MS\kern-0.5pt}\kern0.5pt}

\usepackage{afterpage,float}
\usepackage{enumerate}

\usepackage{amssymb}

\journal{Physics Letters B}

\begin{document}
	
	\begin{frontmatter}

\title{Extended investigation of the twelve-flavor $\beta$-function}

\author[wupi,julich]{Zolt\'{a}n Fodor}

\author[uop]{Kieran Holland}

\author[ucsd]{Julius Kuti\corref{cor1}}
\ead{jkuti@ucsd.edu}

\author[madrid]{D\'{a}niel N\'{o}gr\'{a}di}

\author[wupi]{Chik Him Wong}

\cortext[cor1]{Corresponding author}

\address[wupi]{Department of Physics, University of Wuppertal, 
	Gaussstrasse 20, D-42119, Germany}
\address[julich]{J\"ulich Supercomputing Center, Forschungszentrum, 
	J\"ulich, D-52425 J\"ulich, Germany}
\address[uop]{Department of Physics, University of the Pacific, 
	3601 Pacific Ave, Stockton CA 95211, USA}
\address[ucsd]{Department of Physics 0319, University of California, San Diego, 
	9500 Gilman Drive, La Jolla, CA 92093, USA}
\address[madrid]{Universidad Autonoma, IFT UAM/CSIC and Departamento de Fisica Teorica, 28049 Madrid, Spain}


\begin{abstract}
We report new results from high precision analysis of an important BSM gauge theory with twelve
massless fermion flavors in the fundamental representation of the SU(3) color gauge group.  	
The range of the renormalized gauge coupling is extended  from our earlier work~\cite{Fodor:2016zil}
to probe the existence of an infrared fixed point (IRFP) in the $\beta$-function reported at two different 
locations, originally  in~\cite{Cheng:2014jba} and at a new location in~\cite{Hasenfratz:2016dou}.
We find no evidence for the IRFP of the $\beta$-function in the extended range of the renormalized gauge coupling,
in disagreement with~\cite{Cheng:2014jba,Hasenfratz:2016dou}. New arguments to guard the existence of the IRFP
remain unconvincing~\cite{Hasenfratz:2017mdh}, 
including recent claims of an IRFP with ten massless fermion flavors~\cite{Chiu:2016uui,Chiu:2017kza}
which we also rule out.
Predictions of the 
recently completed 5-loop QCD  $\beta$-function 
for general flavor number are discussed in this context.
\end{abstract}

	
	
	
	

\end{frontmatter}
%
\section{History of the IRFP  with 12 massless fermions}\label{intro}
A conformal infrared fixed point (IRFP) of the $\beta$-function was reported earlier
with critical gauge coupling $g_*^2 \approx 6.2$ and interpreted as conformal behavior
of the much studied BSM gauge theory with twelve massless fermions
in the fundamental representation of the SU(3) color gauge group~\cite{Cheng:2014jba}. 
This result was claimed to confirm the original finding of the IRFP in~\cite{Appelquist:2007hu,Appelquist:2009ty}.
In disagreement with~\cite{Cheng:2014jba,Appelquist:2007hu,Appelquist:2009ty}, the IRFP was  
refuted in~\cite{Fodor:2016zil}. Recently, responding to the negative findings
in~\cite{Fodor:2016zil}, the authors of~\cite{Cheng:2014jba} moved the IRFP to a revised new  location 
$g_*^2 \approx 7$ in~\cite{Hasenfratz:2016dou}. 

The relocation of the IRFP followed the announcement 
of a new IRFP  with ten massless fermion flavors in the fundamental representation of 
the SU(3) color gauge group~\cite{Chiu:2016uui,Chiu:2017kza}.
The claim in~\cite{Chiu:2016uui,Chiu:2017kza}  would imply that 
the theory with twelve flavors must also be conformal and the lower edge of the 
conformal window (CW) of multi-flavor BSM theories with fermions in the fundamental representation 
would be located below ten flavors.  No trace of the reported IRFP with ten flavors was found from
high precision simulations in large volumes~\cite{Kuti:nf10}. Some related predictions of the 
recently completed 5-loop QCD  $\beta$-function for general flavor number will be also discussed in this context.

Results are reported for the $\beta$-function from the analysis of high precision simulations 
in large volumes for $n_f=12$ flavors in two different renormalization schemes and two different implementations
of the gauge field gradient flow on the lattice, 
providing convincing evidence for the non-existence of the IRFP in~\cite{Hasenfratz:2016dou}.
A preview from our forthcoming new publication on the $n_f=10$ $\beta$-function~\cite{Kuti:nf10} is also added
showing evidence for the non-existence of the IRFP published in this theory~\cite{Chiu:2016uui,Chiu:2017kza}.
\section { Scale-dependent  $\beta$-function on the lattice} \label{stepfunction}
The gradient flow based diffusion of the gauge fields on  lattice configurations from 
Hybrid Monte Carlo (HMC) simulations became the method of choice  for studying renormalization 
effects with great accuracy
~\cite{Narayanan:2006rf, Luscher:2010iy, Luscher:2010we,Luscher:2011bx, Lohmayer:2011si}.
In particular, we introduced earlier the gradient flow based
scale-dependent renormalized gauge coupling $g^2(L)$ where the scale is 
set by the linear size $L$ of the finite volume~\cite{Fodor:2012td}. This implementation is based on
the gauge invariant trace of the non-Abelian quadratic field strength,
$E(t) = -\frac{1}{2} {\rm Tr} F_{\mu\nu} F_{\mu\nu}(t)$,
renormalized as a composite operator at gradient flow time $t$ on the gauge configurations
and measured from the discretized lattice implementation, as in~\cite{Luscher:2010iy}.

Following~\cite{Fodor:2012td},  we define 
the one-parameter family of  renormalized non-perturbative gauge couplings 
where the volume-dependent gradient flow time $t(L)$ is set by a
fixed value of  $c = \sqrt{8t}/L$ from the one-parameter family of renormalization schemes.
The renormalized gauge coupling $g^2(L)$ is directly determined from $E(t) = -\frac{1}{2} {\rm Tr} F_{\mu\nu} F_{\mu\nu}(t)$
on the gradient flow of the gauge field at a fixed value of $c$ which defines the renormalization scheme.
The renormalization schemes  $c=0.20$ and $c=0.25$ used  in our work are identical 
to what was used in~\cite{Cheng:2014jba,Hasenfratz:2016dou} including periodic boundary conditions 
on gauge fields and anti-periodic boundary conditions on fermion fields in all four directions of the lattice.

A general method for the scale-dependent renormalized gauge coupling $g^2(L)$  was introduced earlier to probe the
step $\beta$-function, defined as $( g^2(sL) - g^2(L) ) / \log( s^2 )$ for some preset finite scale change $s$
in the linear physical size $L$
of the  four-dimensional volume in the continuum limit of 
lattice discretization~\cite{Luscher:1992an}. 
In our adaptation of the step $\beta$-function staggered lattice fermions are used with stout smearing in the fermion Dirac operator. 
The implementation of the HMC evolution code is
described in~\cite{Fodor:2016zil} together with further details on the lattice 
step function and its continuum limit. Identical procedures are followed here.

%
\section{High precision twelve-flavor analysis in large volumes} \label{nf12}
In the continuum limit, the monotonic function $g^2(L)$ implies in any of the volume-dependent schemes that a selected
value of the renormalized gauge coupling sets the physical size $L$ measured in some particular dimensionful physical unit.
Fixed physical size $L$ on the lattice is equivalent to holding $g^2(L)$ fixed at some selected value as
the lattice spacing $a$ is varied and the fixed physical length $L$ is held constant by the variation of the 
dimensionless linear scale $L/a$ as the bare lattice coupling is tuned without changing the selected fixed value of the
renormalized gauge coupling.
The continuum limit of the $\beta$-function at fixed $g^2(L)$ is obtained by $a^2/L^2\to 0$ extrapolation 
of the residual cut-off dependence detected as powers of $a^2/L^2$ in the step $\beta$-function at 
finite bare gauge couplings $g_0^2$ while the renormalized gauge coupling is held fixed.

In the convention we use, asymptotic freedom in the UV regime corresponds to a positive step $\beta$-function given by
the perturbative loop expansion for small values of the renormalized coupling.
In the infinitesimal derivative limit $s\!\to\!1$ the step  $\beta$-function turns into the conventional one.
It turns out that even at a step size as large as $s=2$ the step $\beta$-function tracks the conventional continuum
$\beta$-function with very good accuracy and we use step $s=2$ throughout the analysis.
If the conventional  $\beta$-function of the theory possesses a conformal fixed point, 
the step $\beta$-function will have a zero at the critical gauge coupling $g_*^2$, independent of the scale $L$. 
Since the renormalized gauge coupling 
$g^2(L)$ is a monotonic function of $L$, the IRFP coupling $g_*^2$ is reached in the $L\to\infty$ limit.

\subsection{SSC gradient flow with c=0.20  renormalization scheme}
Precision tuning of the 
bare gauge coupling $6/g^2_0$ was used exclusively for each step calculation in~\cite{Fodor:2016zil} 
using the $c=0.20$ renormalization scheme with $s=2$  step size. The gauge field gradient flow was driven 
by the same tree-level 
improved Symanzik gauge action which generated the gauge configurations. 
The lattice implementation of the gauge field operator $E(t) = -\frac{1}{2} {\rm Tr} F_{\mu\nu} F_{\mu\nu}(t)$ used the 
clover construction which is known to reduce cut-off effects in the gradient flow~\cite{Luscher:2010iy}.
SSC designates the setup with Symanzik gauge action driving both the gauge field gradient flow 
and the HMC evolution code, together with the 
clover operator implementation of  $E(t)$.
The three target groups A, B, C of the precision tuned run sets tested the IRFP with negative conclusions,
in disagreement with what was reported in~\cite{Cheng:2014jba}.
In this work we target the step $\beta$-function in an extended range of the renormalized gauge coupling 
to cover the interval where the relocated IRFP was recently reported~\cite{Hasenfratz:2016dou}. For consistency check, we added
to the analysis the WSC scheme in section~\ref{label:wsc0p20} where the Symanzik action is replaced 
by the simple Wilson plaquette action to drive the gradient flow on the gauge configurations. 
Only the Wilson plaquette action was used in~\cite{Cheng:2014jba,Hasenfratz:2016dou}
to drive the gradient flow.  Importantly, we also 
test in the new work the influence of $a^4/L^4$ cutoff effects in extrapolations to the continuum $\beta$-function and
extend the analysis to the $c=0.25$ renormalization scheme in section~\ref{label:c0p25}.

In Table~\ref{table:data} results are shown for gauge ensembles 
from the three new target groups D, E, F of precision tuned run sets using 
identical $c=0.2$ renormalization scheme with $s=2$ steps, as before.
\begin{table}[htb] 
     	\begin{center} 
     		\scriptsize
		\begin{tabular}{|c|c|c||c|c||c|c|}
			\hline 
			&\multicolumn{2}{c||}{Target D}&\multicolumn{2}{c||}{Target E}& \multicolumn{2}{c|}{Target F}\\			
			\hline\hline
			L/a& $6/g_0^2$& $g^2$ & $6/g_0^2$ & $g^2$ & $6/g_0^2$ & $g^2$  \\			
			\hline		
			16 & 2.9380 & 6.5972(30)  & & & 2.7838 &6.9855(31)  \\
			\hline 
			32 & 2.9380 &6.4817(136) &  &  & 2.7838 & 6.8237(142) \\
			\hline \hline
			18 &2.9233  &6.5977(34)  & 2.8420 &6.7956(35) & 2.7592  & 6.9834(19)    \\
			\hline 
			36 & 2.9233 &6.5261(166)  &2.8420 & 6.7023(99) & 2.7592 & 6.8461(79)  \\	
			\hline\hline
			20 & 2.9094 &6.5993(53)  & 2.8232 &6.7975(52) & 2.7298 & 6.9870(65)   \\
			\hline 
			40 & 2.9094  &6.5656(67) &2.8232  &6.7423(71) & 2.7298 & 6.8989(75)  \\
			\hline \hline
			24 &2.8932  &6.6006(75)  &2.8006  &6.7910(79)& 2.7000  & 6.9831(71)  \\
			\hline 
			48 & 2.8932&6.6229(125) & 2.8006&6.8083(94)  & 2.7000 & 6.9764(85) \\	
			\hline\hline		
			28 &2.8817  &6.6012(41)   &  &  & 2.6884  & 6.9820(49)  \\
			\hline 
			56 & 2.8817 & 6.6898(155)  & &  & 2.6884 & 7.0133(111)  \\	
			\hline\hline								
		\end{tabular} 
    	\end{center} 	
		\caption{ With previously  tuned bare gauge couplings $ g_0^2$, the final 26 precision tuned runs are tabulated 
			with 13 tuned runs and 13 paired $s=2$ steps.  The D, E, F run sets target  $g^2$ approximately at 6.6, 6.8, 7.0 respectively. }
	    \label{table:data} 
\end{table}

\begin{figure}[htb!]
	\centering
	\includegraphics[width=0.8\linewidth]{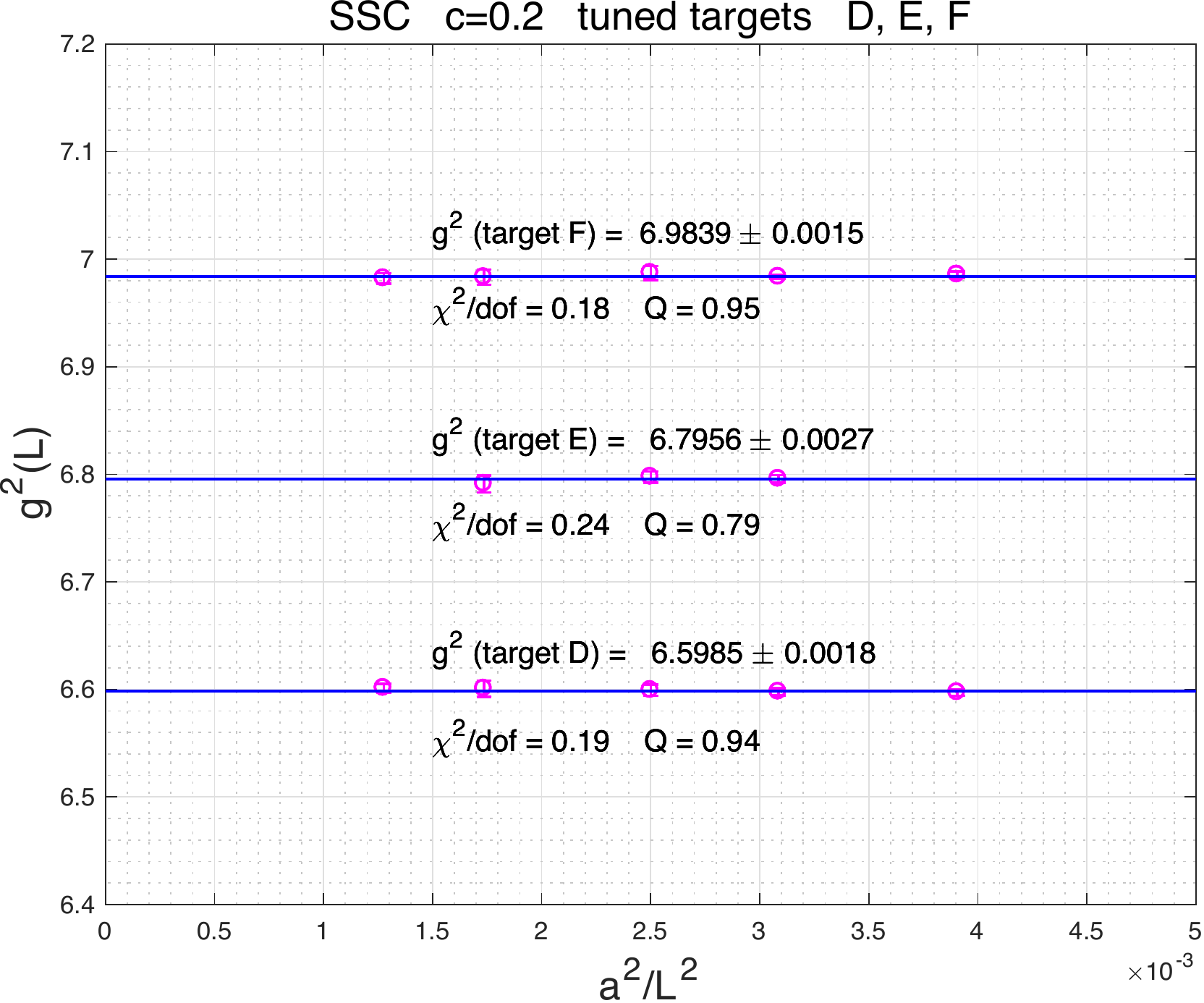}	
	\caption{\label{fig:tuning}  The statistical significance of precision tuning to three targeted gauge couplings 
		D, E, F is shown by fitting a constant to each $g^2$ at the lower $L/a$ values of each step.}
\end{figure}

The 26 runs were grouped into pairs for each step where the lower $L/a$ value was 
precisely tuned to the target value of the renormalized gauge coupling. The higher $L/a$  at the doubled physical
size determined the step $\beta$-function at finite lattice spacing.
Precision tuning for  $g_0^2$ of the 13 steps of the three targets eliminated
the systematic  uncertainty in the step $\beta$-function from model-dependent interpolation in the bare gauge coupling. 
Figure~\ref{fig:tuning} shows the remarkable accuracy of tuning for the three targets on the per mille accuracy level, 
with similar accuracy level of the renormalized $g^2$ entries in Table~\ref{table:data}. 

The statistical analysis of the renormalized gauge coupling and step $\beta$-function of the precision tuned runs followed~\cite{Wolff:2003sm} 
and used similar software. 
For each run, extended in length 
between 5,000 and 20,000 time units of molecular dynamics time, autocorrelation times were measured
in two independent ways, using estimates from the autocorrelation function 
and from the Jackknife blocking procedure. Errors on the renormalized couplings were
consistent from the two procedures and the one from autocorrelation functions is listed in Table~\ref{table:data}.
Each run went through thermalization segments which were not included in the analysis.
For detection of residual thermalization effects the replica method of~\cite{Wolff:2003sm}  
was used in the analysis. All 26 runs passed Q value tests when mean values and statistical errors of the replica segments 
were compared for thermal and other variations. 

\begin{figure}[hbt!]
	\centering
	\includegraphics[width=0.8\linewidth]{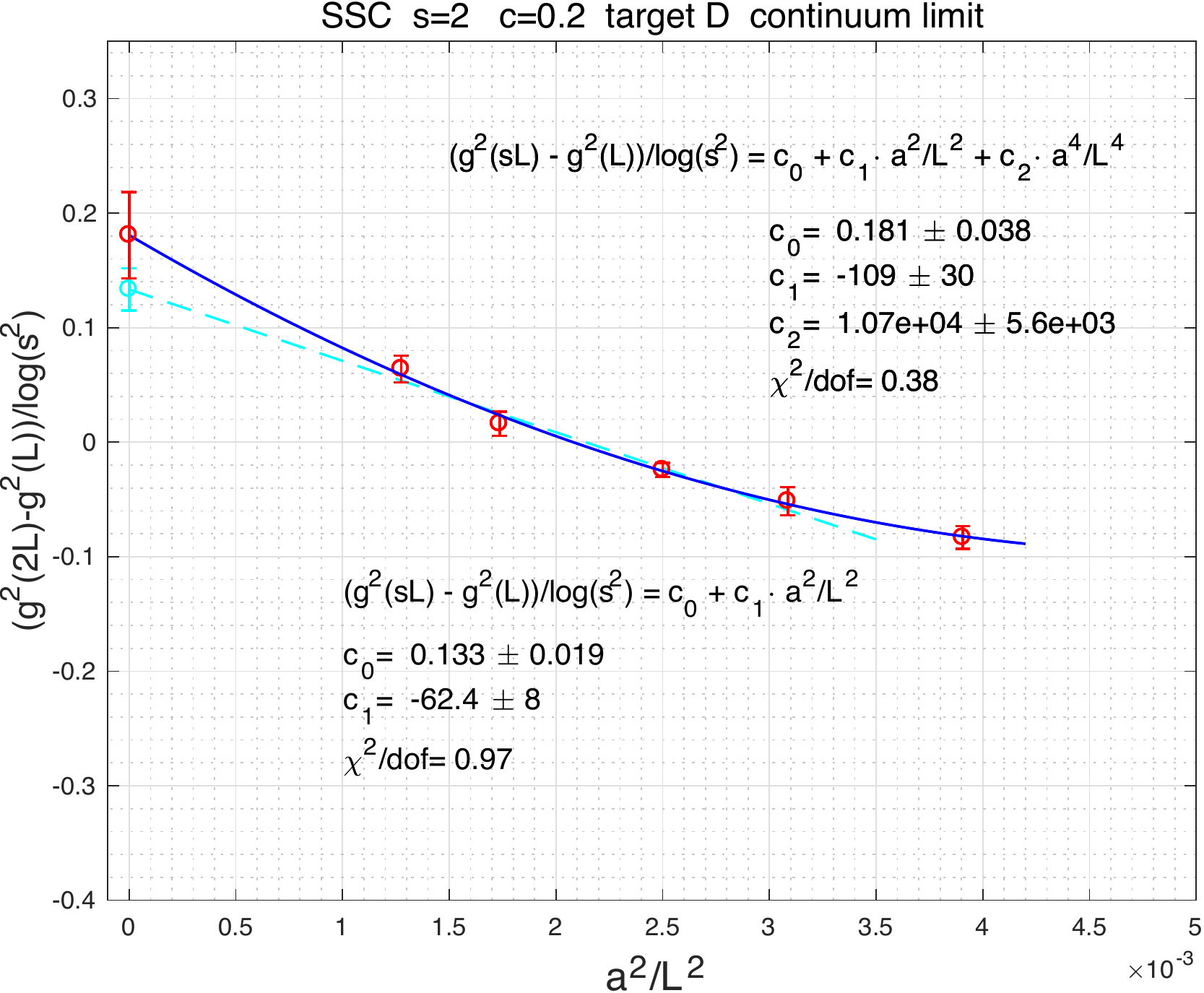}\\
	\vskip 0.1in
	\includegraphics[width=0.8\linewidth]{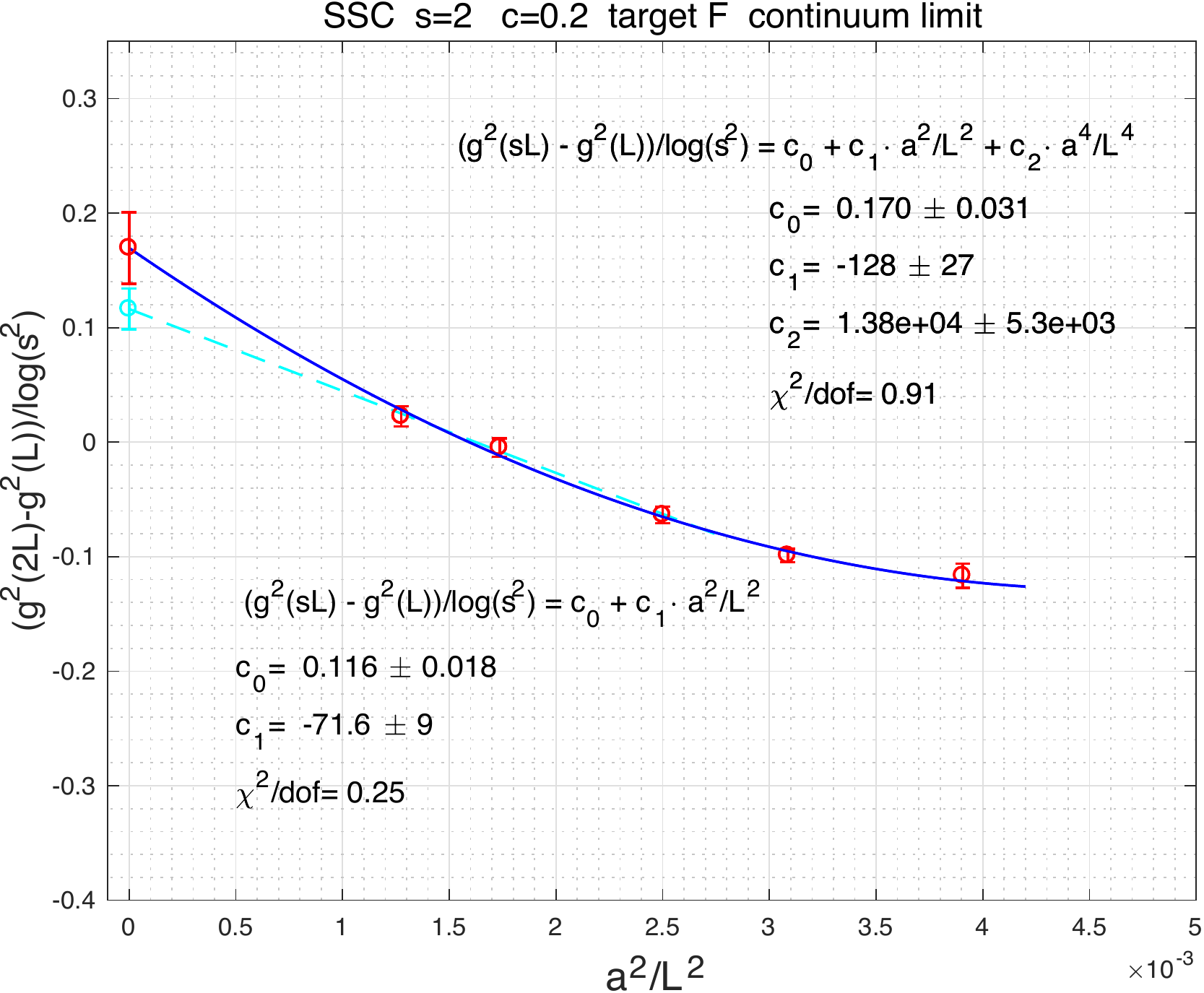}	
	
	\caption{\label{fig:targets} 
		The consistency of extrapolations to the continuum $\beta$-function is illustrated
		comparing linear 3-point fits and quadratic 5-point fits in the $a^2/L^2$ variable for target D (top) and target F (bottom).}
\end{figure}

The leading cutoff effects in the step function at finite bare coupling $g_0^2$ appear at $a^2/L^2$ order with $a^4/L^4$ 
and higher order corrections. In our earlier work~\cite{Fodor:2016zil} the linear dependence on $a^2/L^2$ was detected and fitted 
without including higher order corrections in reaching the continuum limit of the $\beta$-function. 
The high precision of the data allows for testing the $a^4/L^4$  correction term leading to small increases of
the continuum $\beta$-function within one standard deviation in the SSC setup. This is  
illustrated in  Figure~\ref{fig:targets} for targets D and F.

\begin{figure}[thb] %
	\centering
	\includegraphics[width=0.8\linewidth,clip]{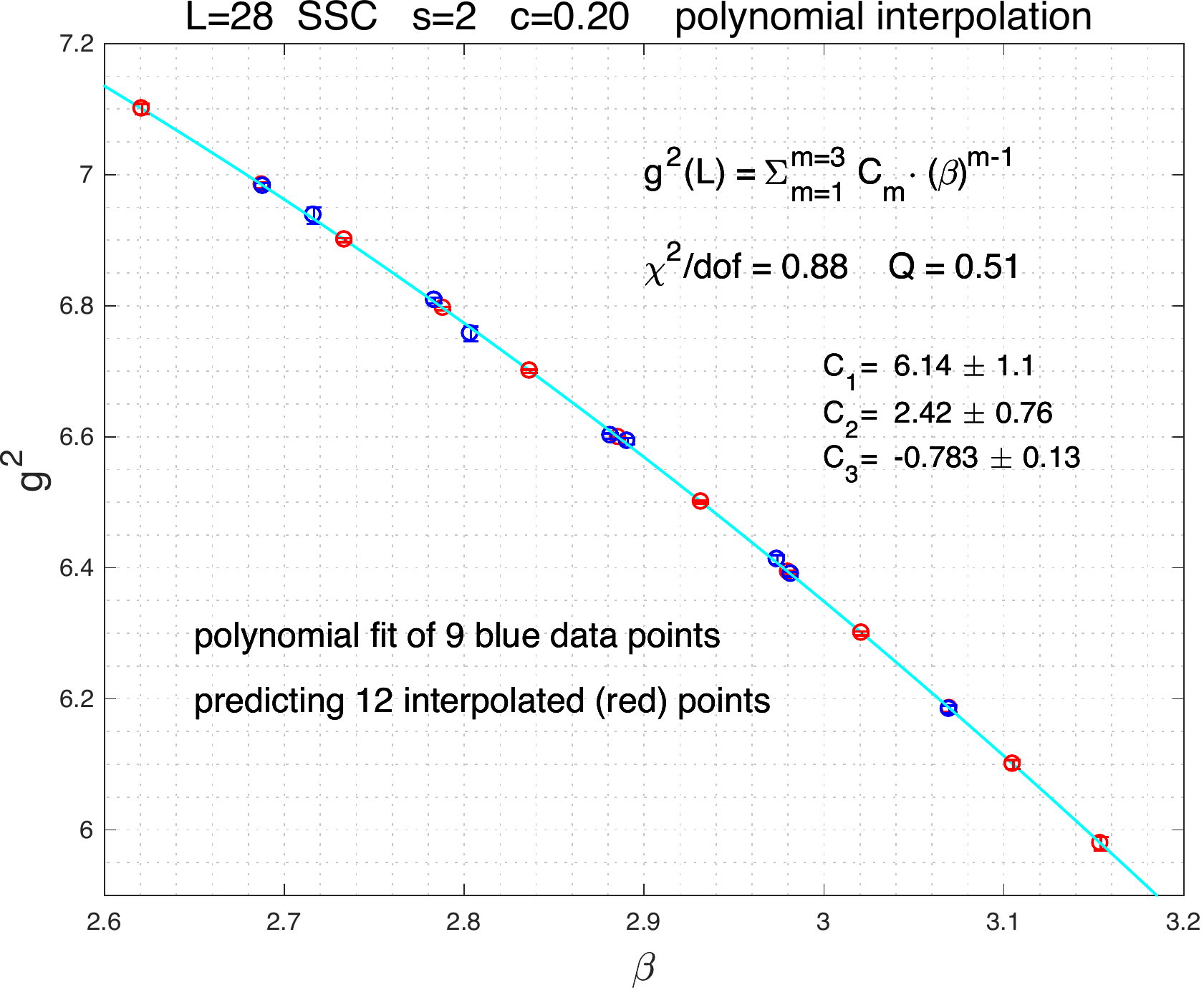}\\
	\vskip 0.1in
	\includegraphics[width=0.8\linewidth,clip]{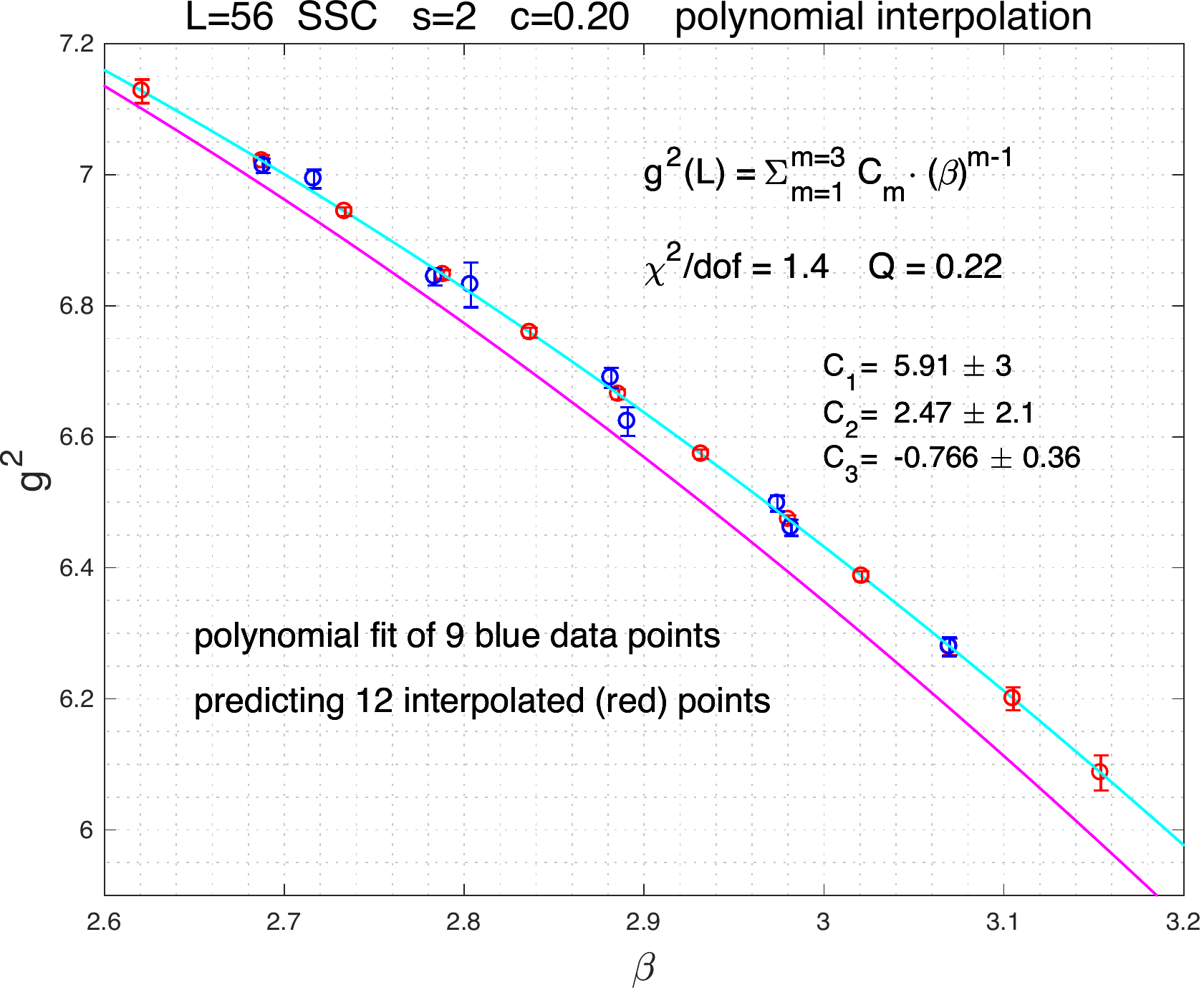}
	\caption{\footnotesize Polynomial interpolation is shown for 12 targeted gauge couplings (red points) using 9 inputs (blue points) from 
	6 runs of precision tuning and from additional auxiliary runs used in the tuning procedure. The largest $28\rightarrow 56$ step is shown 
    with the magenta line marking the fit of the L=28 data (top) overlayed on the L=56 plot (bottom). The  step $\beta$-function is positive
    throughout the targeted range.} 
	\label{fig:L28}
\end{figure}

In addition to targeted precision tuning all the runs from the 6 targets A-F can be combined with additional trial runs from the tuning procedure 
into a new extended analysis which can project renormalized gauge couplings and step functions at any location 
in the bracketed range by using simple polynomial interpolation. 
Two samples of all the high quality polynomial interpolations are shown in Figure~\ref{fig:L28}. This procedure allowed us to include the WSC analysis 
and the $c=0.25$ renormalization scheme in the new work. 

The results shown in Figure~\ref{fig:L28} are quite remarkable. The largest $28\rightarrow 56$ step  
probes the smallest value of $a^2/L^2$  with a positive step $\beta$-function over the 
whole bracketed $g^2$  range, 
incompatible with a zero in the $\beta$-function as first evidence against the existence of the IRFP.

\begin{figure}[thb] 
	\centering
	\includegraphics[width=0.8\linewidth,clip]{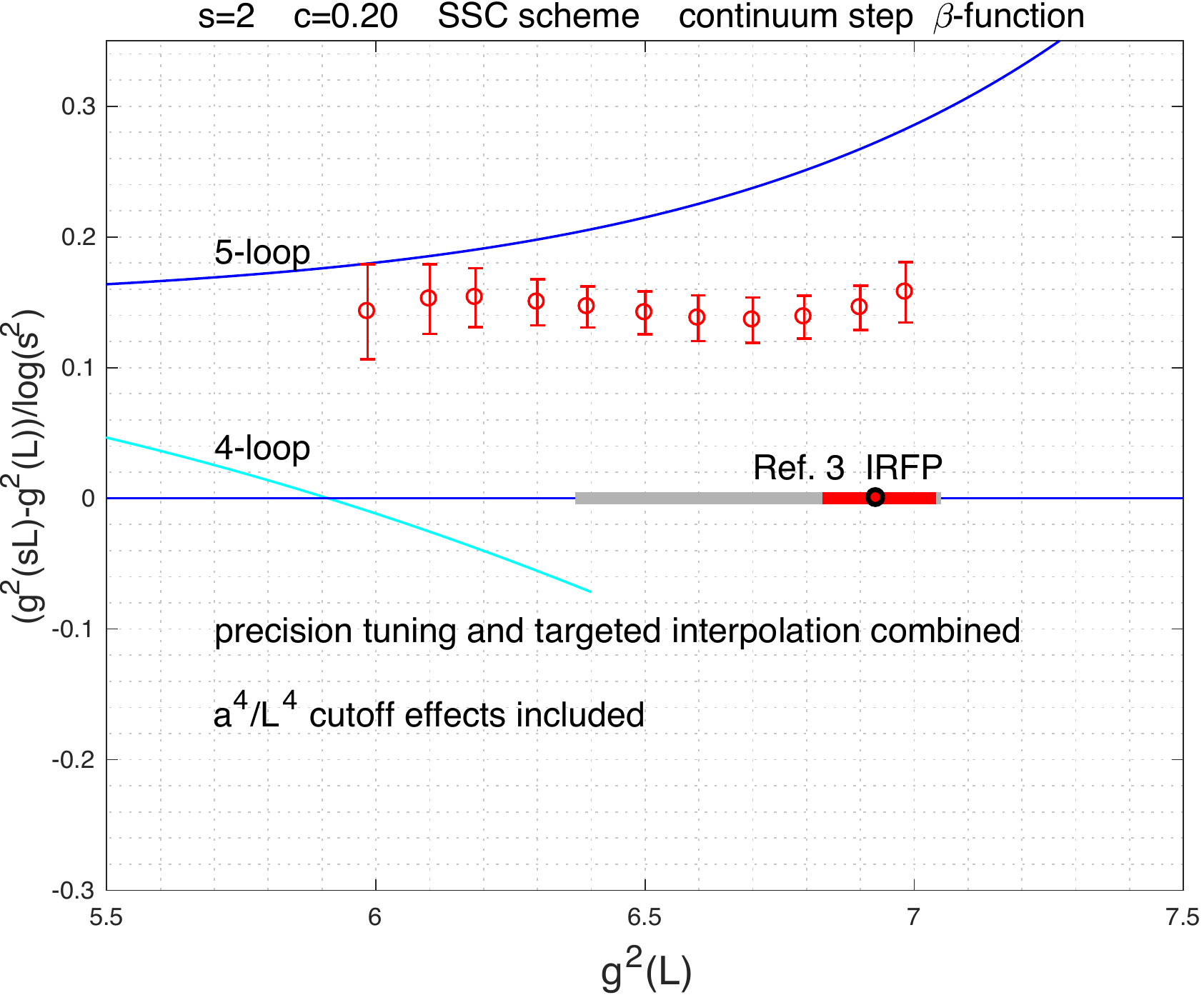}\\
	\vskip 0.1in
	\includegraphics[width=0.8\linewidth,clip]{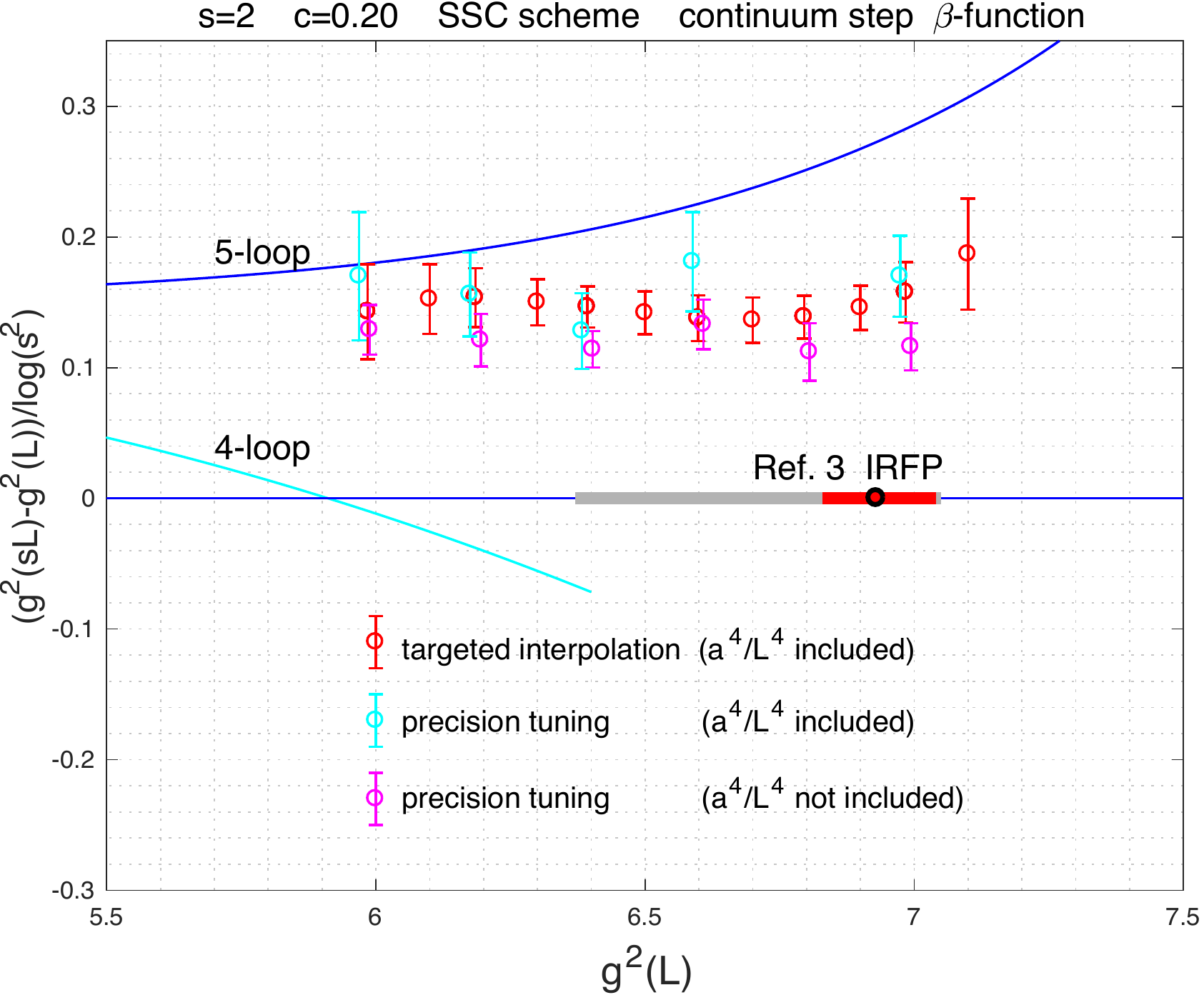}
	\caption{\footnotesize In the top panel results are shown using simple polynomial interpolation 
             for 6 combined precision tuned target runs and 3 additional auxiliary runs. Fits in extrapolation of the steps 
             include the $a^4/L^4$ cutoff effects as shown in Figure~\ref{fig:w20}. 
            The IRFP with red bar for statistical error and grey bar for systematic estimate
            is from~\cite{Hasenfratz:2016dou}.                 
            The lower panel, with slightly shifted data in cyan and magenta colors for visibility,
             compares results from precision tuning and the combined method with an extrapolated point 
            predicted slightly outside the bracketed interpolation range.        
            Linear 3-point fits in $a^2/L^2$ were used for precision tuned targets A and B (4-point fits before in~\cite{Fodor:2016zil} ).
            Both plots display the 4-loop and recent 5-loop results referenced in the main text.}	
	\label{fig:beta20}
\end{figure}

The main results of the SSC analysis of the $c=0.20$ renormalization scheme with step size $s=2$ are shown in Figure~\ref{fig:beta20}.
The top panel of the figure is the outcome of the analysis from the combined use of the precision tuned 6 target runs and auxiliary runs 
using simple polynomial interpolation for 11 predictions which include the 6 locations of precision tuning for consistency and 
enhanced accuracy.
The lower panel in Figure~\ref{fig:beta20} compares results from precision tuning with the combined method showing remarkable consistency.
Both the upper and lower panels display the 4-loop and recent 5-loop results of the 
continuum $\beta$-function in the $g_{\msbar} ^2$ scheme~\cite{Baikov:2016tgj,Ryttov:2016asb,Ryttov:2017kmx,Luthe:2017ttc,Herzog:2017ohr}.
Extrapolation of the steps from finite bare couplings $g_0^2$ to the continuum $\beta$-function 
includes the $a^4/L^4$ cutoff effects with typical fits shown in Figure~\ref{fig:w20}. 

In the 4-loop approximation the $n_f=12$ theory has an IRFP which disappears in the 5-loop  approximation. 
The 5-loop $\beta$-function predicts the lower edge of the conformal window  between $n_f=12$ and $n_f=13$.
It will require further investigation to understand the potential significance of the apparent consistency between 
the 5-loop $\beta$-function and our simulation results in different 
renormalization schemes. 
The authors of~\cite{Cheng:2014jba,Hasenfratz:2016dou,Hasenfratz:2017mdh} prefer to show only the 4-loop $\beta$-function
which exhibits a zero at the $g^2$ location close to where the original IRFP was 
published in~\cite{Cheng:2014jba} before being relocated to $g^2\approx 7$ in the 
$c=0.20$ renormalization scheme~\cite{Hasenfratz:2016dou,Hasenfratz:2017mdh}. 
The 4-loop zero in the $\beta$-function is inconsistent with the new 5-loop results.
Independently, and non-perturbatively, the reported IRFP is ruled out in our new analysis of the extended data set with
overwhelming statistical significance as illustrated in Figure~\ref{fig:beta20}.

\subsection{WSC gradient flow with c=0.20  renormalization scheme} \label{label:wsc0p20}
\begin{figure}[bht!] %
	\centering
	\includegraphics[width=0.70\linewidth,clip]{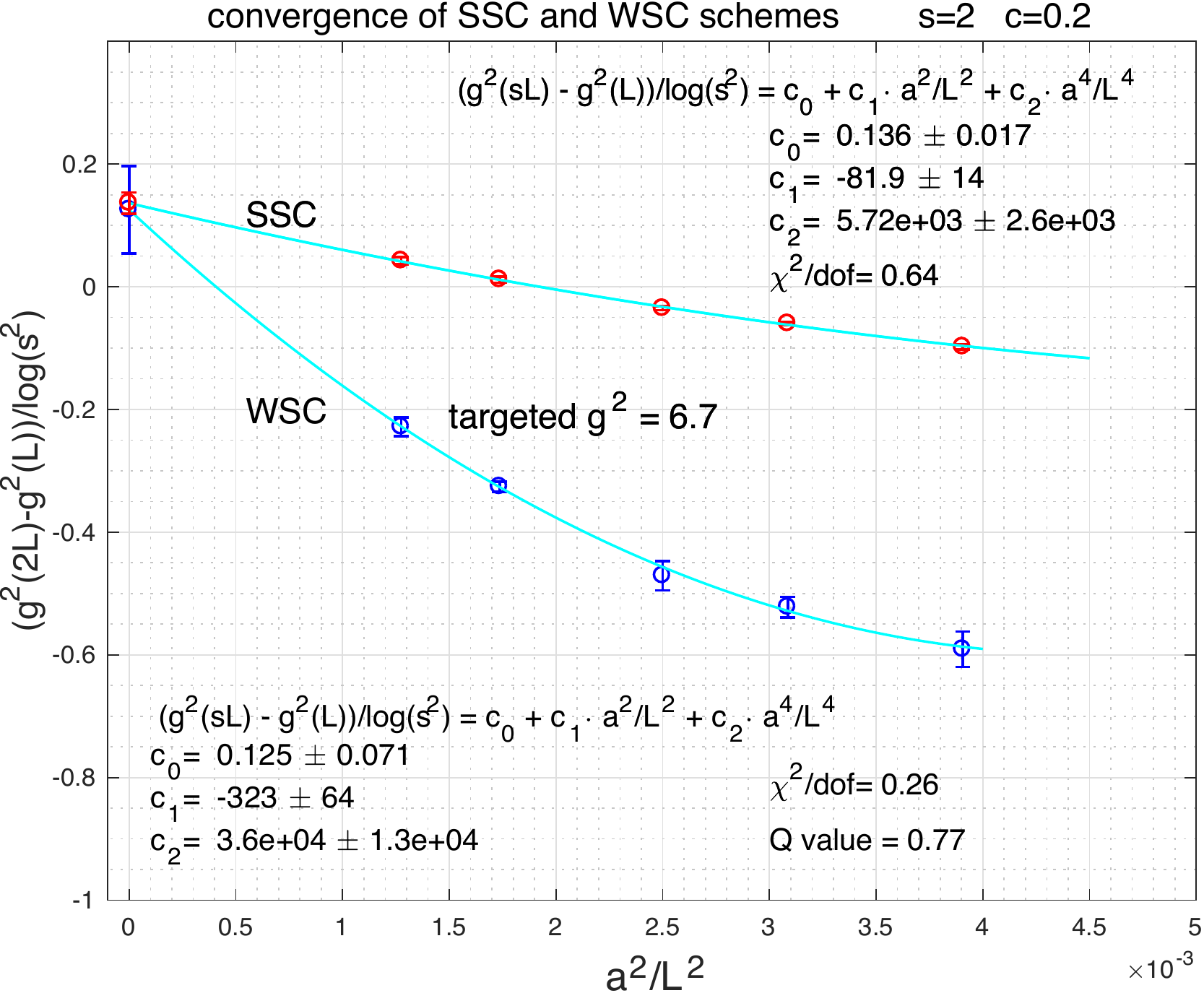}\\
	\vskip 0.07in
	\includegraphics[width=0.70\linewidth,clip]{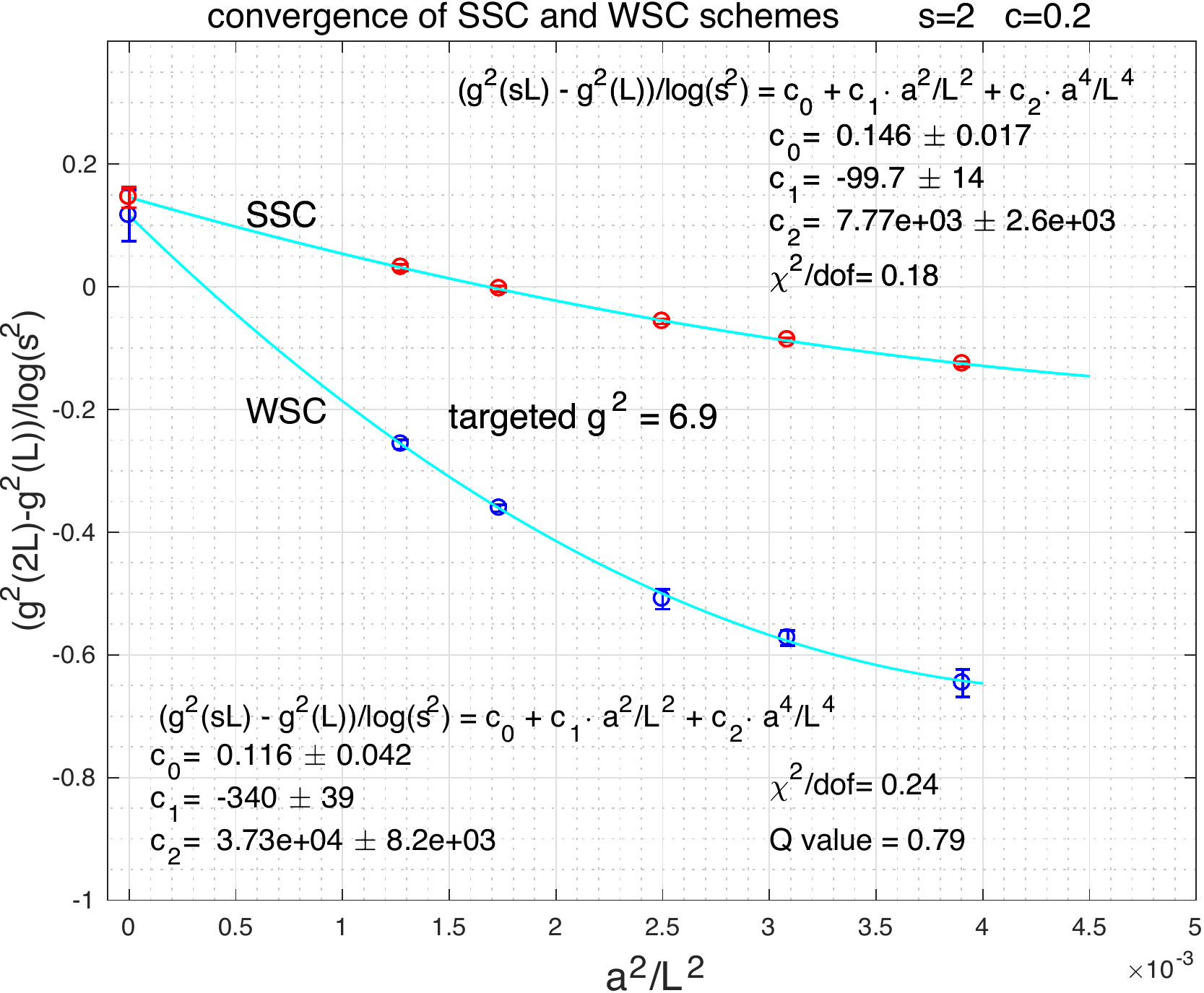}\\
	\vskip 0.07in
	\includegraphics[width=0.70\linewidth,clip]{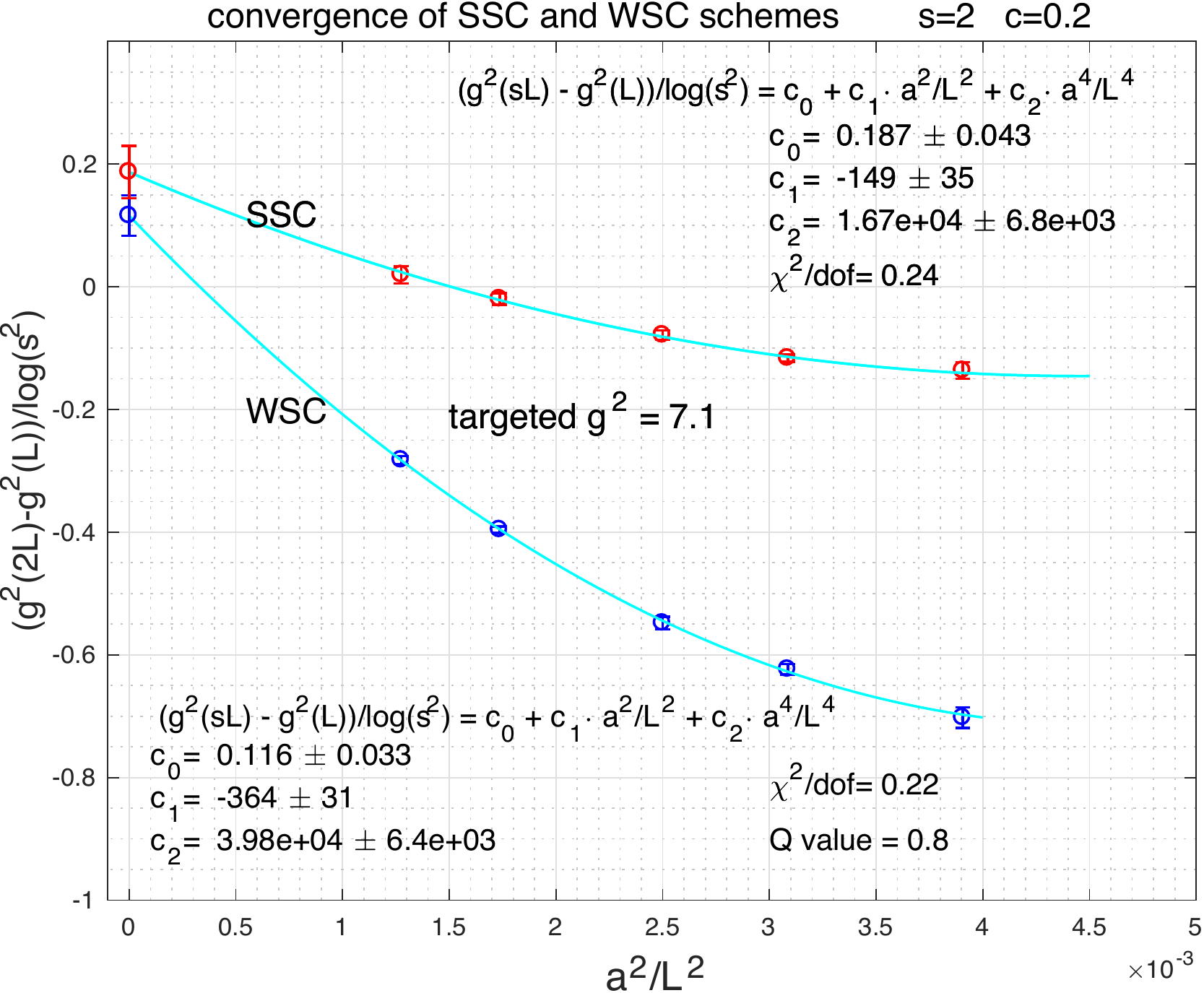}
	\caption{ The fitting procedure is shown for SSC and WSC analyses where Symanzik action
		and Wilson action are driving the gauge field gradient flow respectively. The analysis of the extrapolation
		to the continuum $\beta$-function includes in both cases the $a^2/L^2$ and the $a^4/L^4$ terms in the 
		fitting procedure. }
	\label{fig:w20}
\end{figure}

The WSC setup in our analysis designates the replacement of the tree-level improved Symanzik action
by the simple Wilson plaquette action to drive the gradient flow on the gauge configurations
which were generated by the Symanzik action in the HMC evolution code.
The $a^4/L^4$ cutoff contamination plays a critically important role 
in establishing consistency between the WSC scheme and the much less cutoff contaminated SSC scheme
in reaching the correct continuum $\beta$-function. Since only the Wilson plaquette action was used for 
the gradient flow in~\cite{Cheng:2014jba,Hasenfratz:2016dou}, the WSC analysis is useful  for completeness 
and consistency checks.
Figure~\ref{fig:w20} shows the consistency of the two gradient flows converging to the same continuum 
$\beta$-function in the $c=0.20$ renormalization scheme with step size $s=2$. 
The analysis of extrapolations 
to the continuum $\beta$-function included both the $a^2/L^2$ term and the $a^4/L^4$ term in the fitting procedure.
It is quite remarkable that consistency is clearly established although the cutoff effects in the WSC gradient flow 
are almost an order of magnitude larger than in the SSC gradient flow. 
\vskip 0.1in

\begin{figure}[thb] %
	\centering
	\includegraphics[width=0.8\linewidth,clip]{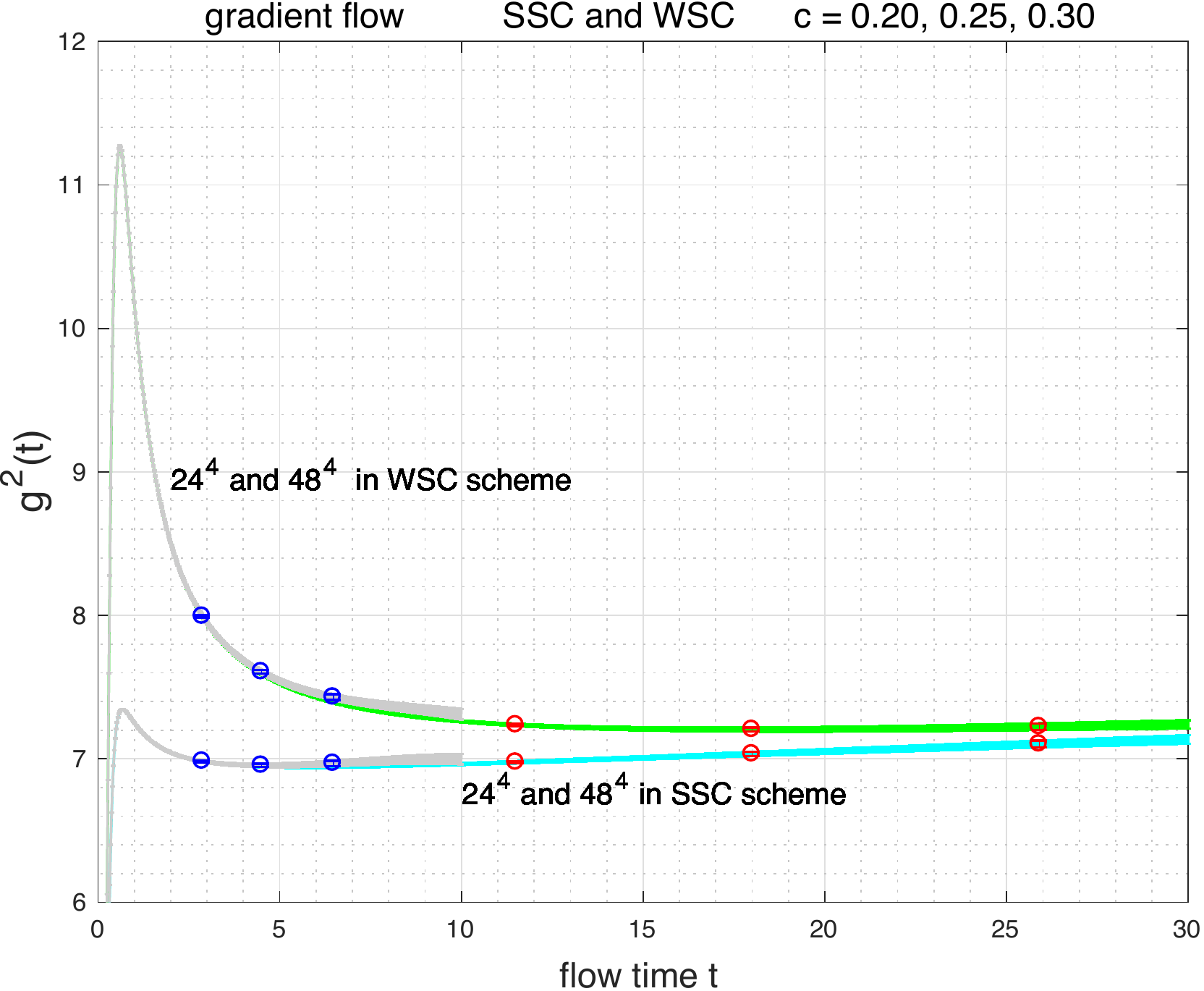}\\
	\vskip 0.1in
	\includegraphics[width=0.8\linewidth,clip]{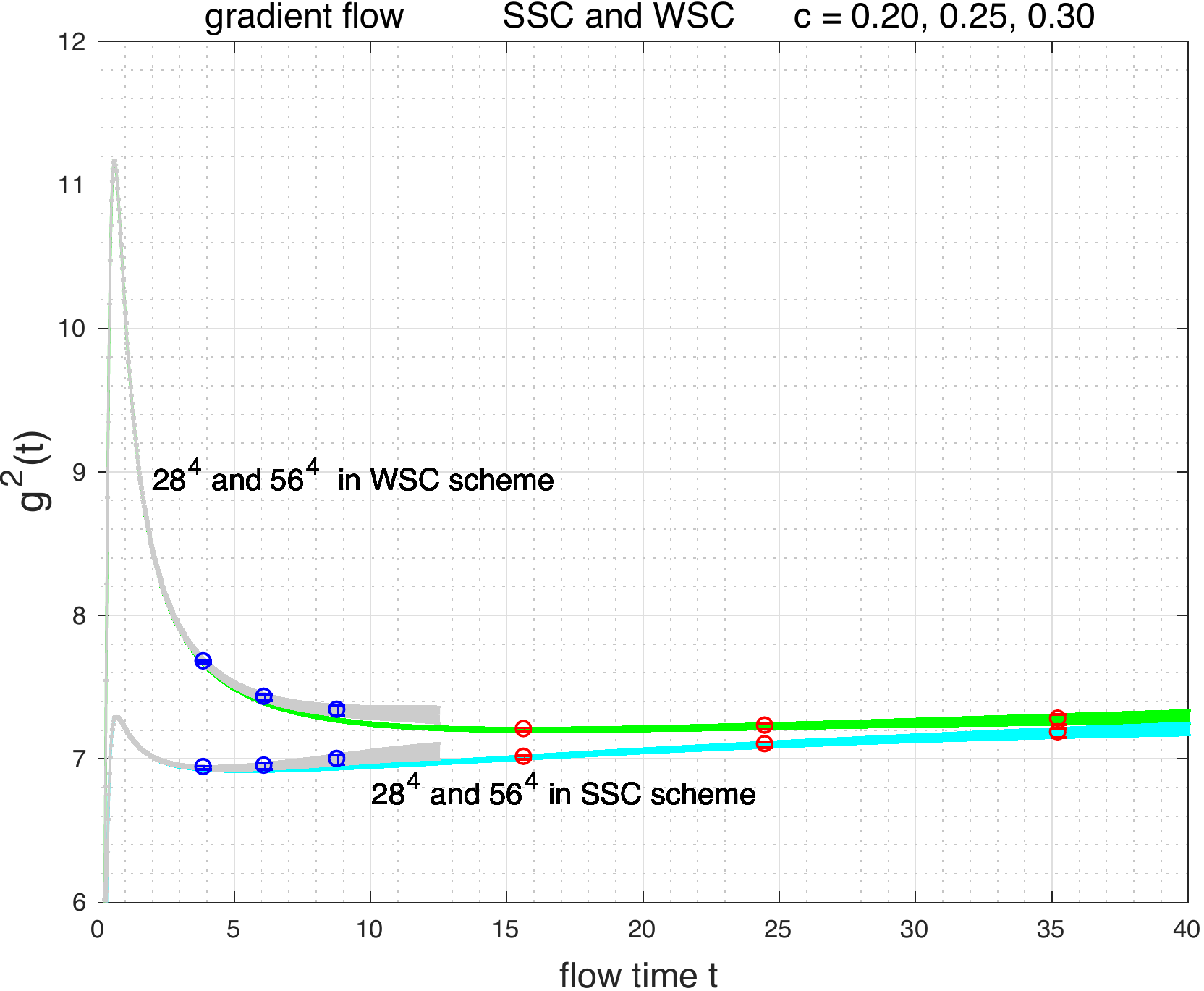}
	\caption{ For renormalization schemes $c=0.20/0.25/0.30$, the locations of the paired points of the
	step $\beta$-functions are shown separately for  Wilson and Symanzik action driving the gradient flow.
	Grey color shades the flow of the smaller $L$ of the pair with blue data points, green and cyan colors shade the larger $L$ with red data points. 
	The top panel is the $24\rightarrow 48$ step, the lower panel shows the $28\rightarrow 56$ step. The target F set of Table~\ref{table:data}
    was used at all 3 values of $c$ for SSC and WSC.}
	\label{fig:L48}
\end{figure}

\noindent{\em The origin of the large WSC cutoff effects:}
The lattice implementation of composite operators gets renormalized along the gradient flow as a function of flow time $t$. 
Figure~\ref{fig:L48} shows the renormalization of the clover lattice implementation of the composite operator 
$E(t) = -\frac{1}{2} {\rm Tr} F_{\mu\nu} F_{\mu\nu}(t)$ along the gradient flow, proportional to $g^2(t)$ we target. 

The renormalization of the operator $E(t)$  goes  through transient effects at short flow times 
before the cutoff effects get sufficiently renormalized. These transient effects are almost an order of magnitude larger 
along the Wilson flow for the same clover lattice implementation as used in the Symanzik flow. In our WSC approach,
much larger stepped volumes would be needed to match the  smaller cutoff effects in 
extrapolation of the SSC approach to the 
continuum $\beta$-function. This is clearly demonstrated in Figure~\ref{fig:w20} and in Figure~\ref{fig:L48}. 
We do not see how the stronger cutoff dependence of the Wilson flow allows one to ignore the $a^4/L^4$ effects
of small lattices when  gauge and fermion 
actions different from ours are used in generating configurations in the simulations, like in~\cite{Chiu:2016uui,Chiu:2017kza,Hasenfratz:2017mdh}.

\subsection{The c=0.25 renormalization scheme} \label{label:c0p25}
\begin{figure}[bh!] 
	\centering
	\includegraphics[width=0.7\linewidth,clip]{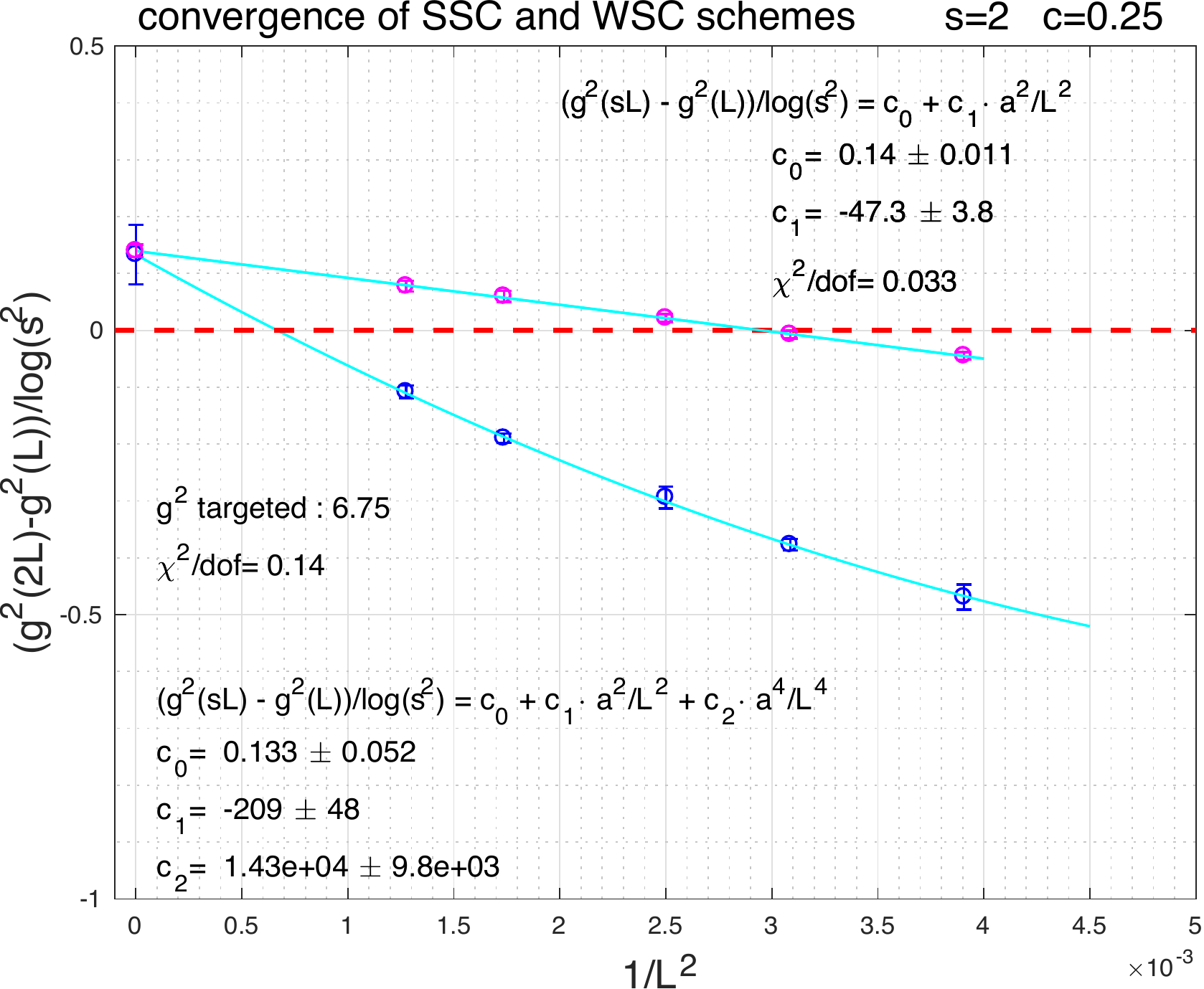}\\
	\includegraphics[width=0.7\linewidth,clip]{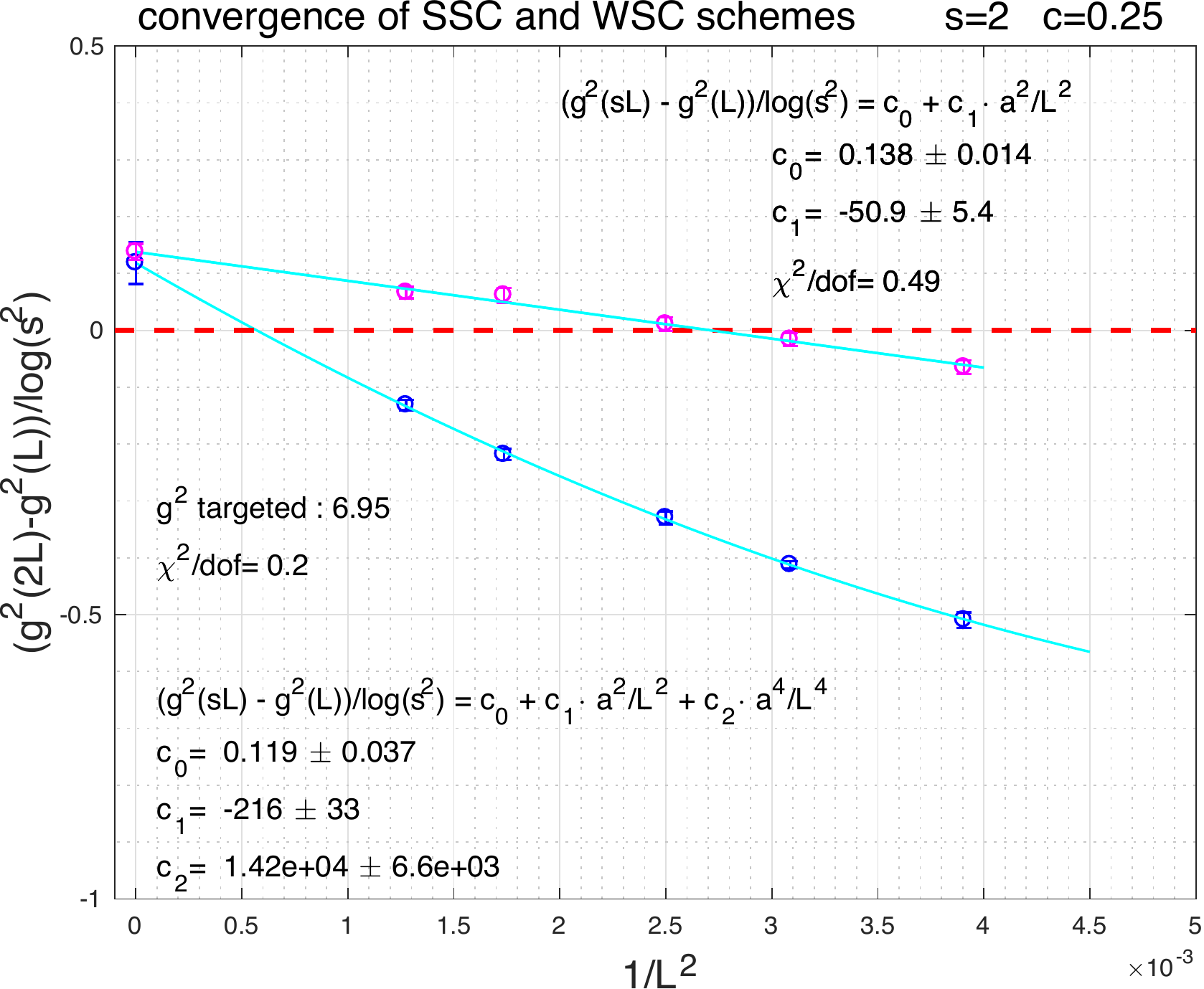}\\
	\includegraphics[width=0.7\linewidth,clip]{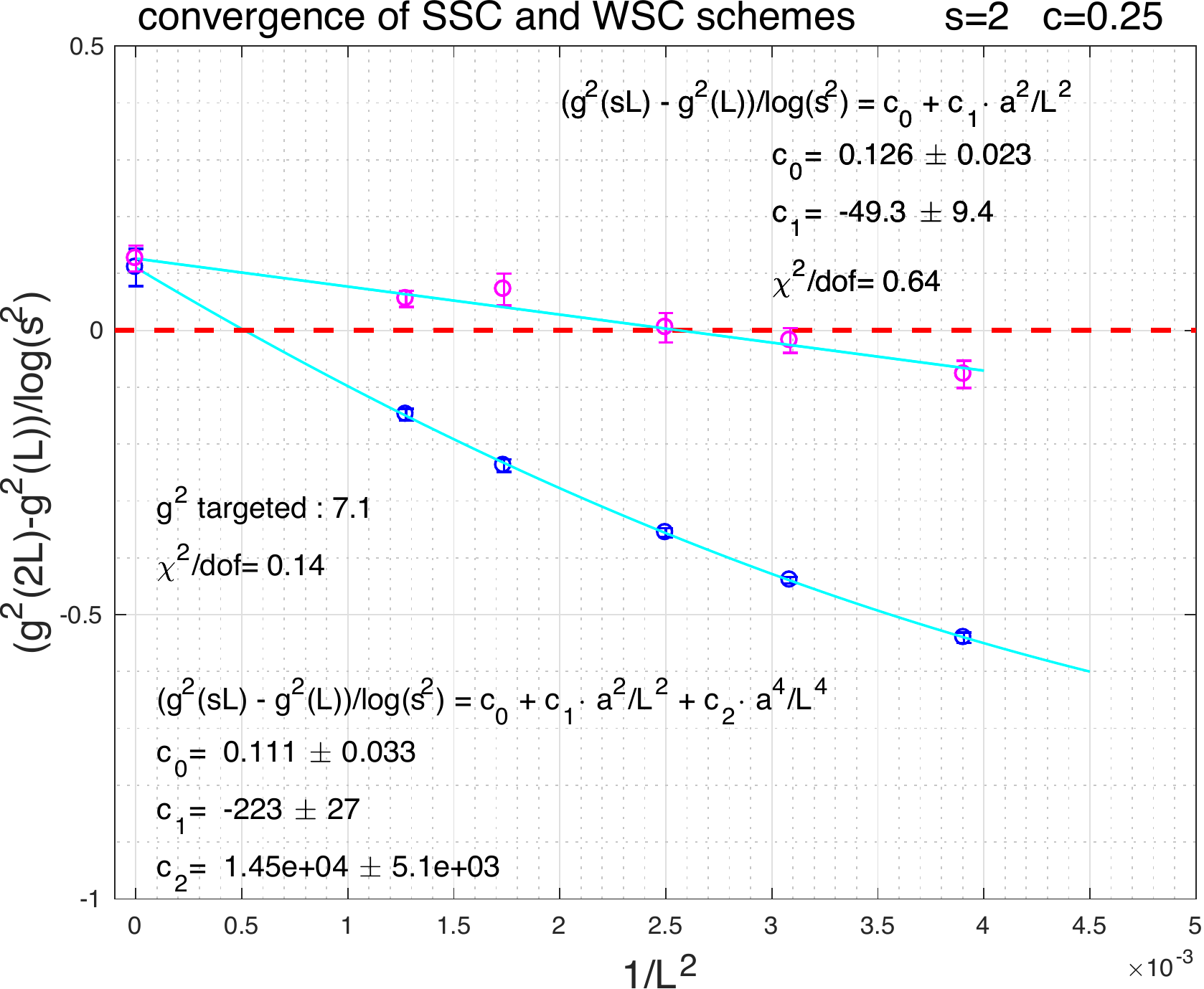}
	\caption{ The fitting procedure is shown in the $c=0.25$ renormalization scheme 
		for SSC and WSC analyses where Symanzik action
		and Wilson action are driving the gauge field gradient flow respectively. The analysis of the extrapolation
		to the continuum $\beta$-function includes $a^2/L^2$ and $a^4/L^4$ terms in the WSC
		fitting procedure. }
	\label{fig:c0p25}
\end{figure}

\begin{figure}[thb] 
	\centering
	\includegraphics[width=0.8\linewidth,clip]{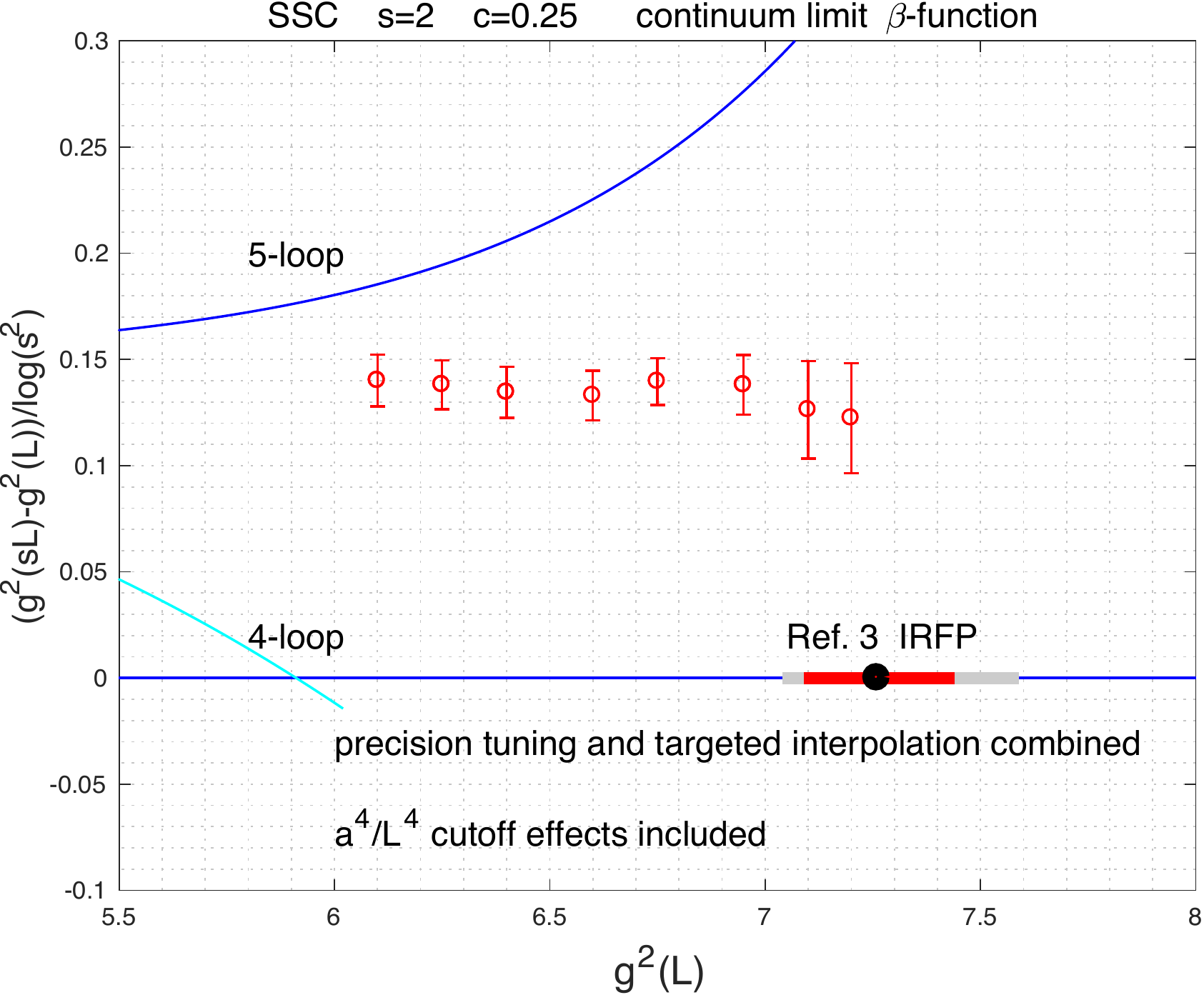}
	\caption{ Results are shown for SSC $c=0.25$ with step size $s=2$ using simple polynomial interpolation 
		for 6 combined precision tuned target runs and 3 additional auxiliary runs. Linear fits are used in the SSC analysis of the steps 
		and the WSC analysis for cross checks  includes $a^4/L^4$ cutoff effects as shown in Figure~\ref{fig:c0p25}. The location of the IRFP in the
		$c=0.25$ scheme of~\cite{Hasenfratz:2016dou} is somewhat shifted to the right from the location of the IRFP  in the
		$c=0.20$ scheme as displayed in this figure and in Figure~\ref{fig:beta20}.
        The 4-loop and recent 5-loop results for the $\beta$-function are referenced in the main text.} 
	\label{fig:beta0p25}
	\vskip -0.1in
\end{figure}

We also analyzed our data set in the $c=0.25$ renormalization scheme which is further away from the $c=0$ infinite volume scheme and
less sensitive to cutoff effects. A subset of the full $c=0.25$ analysis is shown in Figure~\ref{fig:c0p25}.
We combine again all the runs from precision tuning of the 6 targets A-F with additional trial runs from the tuning procedure
into a new extended analysis with SSC and WSC gradient flow and the $c=0.25$ renormalization scheme at step size $s=2$.
In this new analysis we can predict again renormalized 
gauge couplings and step functions at any location in the bracketed range by using simple polynomial interpolation. 
 Although the cutoff effect of the renormalization is much bigger when the Wilson action
is driving the gradient flow, the two consistently converge to the same continuum $\beta$-function.
The $\beta$-function in Figure~\ref{fig:beta0p25} is consistent with the $c=0.20$
renormalization scheme and without any trace of the IRFP reported in~\cite{Cheng:2014jba,Hasenfratz:2016dou}     
for the $c=0.25$ renormalization scheme.

\section{Ten-flavor preview with conclusions}
%
%
We explored 
SSC and WSC gradient flows in the $c=0.20$ and $c=0.25$ renormalization schemes for the determination of the 
continuum $\beta$-function with twelve flavors of massless fermions. Consistent results from our 
high precision simulations in large volumes do not show any infrared fixed point
reported at several locations  in~\cite{Cheng:2014jba,Hasenfratz:2016dou,Hasenfratz:2017mdh}. This disagreement requires
resolution and closure. 
Arguments were presented in~\cite{Hasenfratz:2017mdh} that the different results should not be viewed as disagreements 
in the simulations and in their analysis
but as evidence for the violation of universality
in the staggered formulation. 
Supporting this argument, three examples were invoked  in~\cite{Hasenfratz:2017mdh}:
\vskip 0.1in

\begin{enumerate}[(a)]
	
\item showing conformal fixed point structures in 3D statistical models with several fixed points,
	
\item disagreement between sextet $\beta$-functions
using Wilson fermions~\cite{Hasenfratz:2015ssa} and rooted staggered fermions~\cite{Fodor:2015zna},

\item new results finding an IRFP with ten 
flavors of massless domain wall fermions  implies conformal behavior for twelve flavors as well.

\end{enumerate}

\newpage
\noindent  We note our disagreements in response:
\vskip 0.1in

\begin{figure}[h!] %
	\centering
	\includegraphics[width=0.8\linewidth,clip]{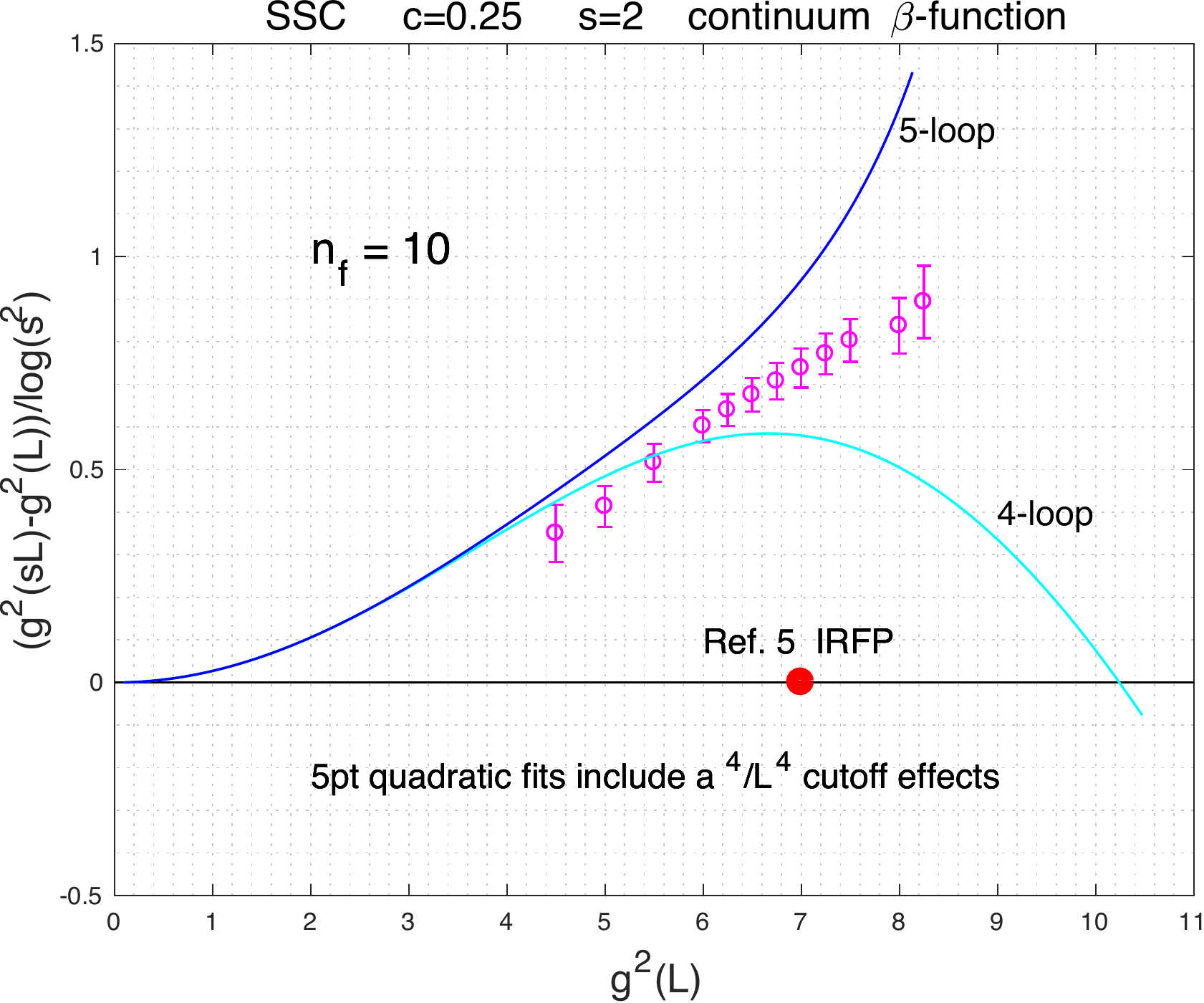}
	\caption{ 
	   The IRFP as marked in the plot is taken from~\cite{Chiu:2016uui,Chiu:2017kza},  in complete 	
	   disagreement with our analysis.}
	\label{fig:nf10}
\end{figure}

\begin{enumerate}[(a)]

\item Since staggered fermions at $n_f=12$ are built on a UV fixed point at zero gauge coupling,
relevant or marginal operators, like in the examples of the 3D statistical models in~\cite{Hasenfratz:2017mdh}, 
cannot be added to the staggered lattice fermion action which has correct locality and universality properties. 
The explicit construction is well-known in the literature. Besides, the controversy 
between~\cite{Cheng:2014jba,Hasenfratz:2016dou,Hasenfratz:2017mdh} and our work exists for the staggered 
formulation itself.

\item The theoretical framework for the rooting procedure, without universality violation
when the sextet gauge coupling is targeted in fixed physical volume,
was explained in~\cite{Fodor:2015zna}.  It has not been challenged since in any follow up to 
the original criticism~\cite{Hasenfratz:2015ssa}. 

\item We recently completed the comprehensive analysis of the theory with ten 
massless fermion flavors in the fundamental representation of the SU(3) color gauge group. 
An important result is shown in Figure~\ref{fig:nf10} for the preview of the $\beta$-function with SSC analysis of
the gradient flow in the $c=0.25$ renormalization scheme using step size $s=2$~\cite{Kuti:nf10}.
Every detail of the ten-flavor analysis followed closely the procedure we implemented and used 
here for the analysis of the twelve-flavor model. 
Since rooting was used for ten flavors with staggered fermions, we tested the validity of the theoretical 
argument presented in~\cite{Fodor:2015zna}. The Dirac spectrum was closely analyzed in the runs 
and the expected behavior of the quartet structure was confirmed.
\end{enumerate}

In conclusion, we find unacceptable the resolution of the problem as conjectured in~\cite{Hasenfratz:2017mdh}.
\newpage
\section*{Acknowledgments}
We acknowledge support by the DOE under grant DE-SC0009919, by the NSF under grants 1318220 and 1620845, 
by OTKA under the grant OTKA-NF-104034, and by the Deutsche
Forschungsgemeinschaft grant SFB-TR 55. Computational resources were provided by the DOE INCITE program
on the ALCF BG/Q platform, USQCD at Fermilab, 
by the University of Wuppertal, by Juelich Supercomputing Center on Juqueen
and by the Institute for Theoretical Physics, Eotvos University. We are grateful to Szabolcs Borsanyi for his code 
development for the BG/Q platform. We are also 
grateful to Sandor Katz and Kalman Szabo for their CUDA code development.

\bibliographystyle{elsarticle-num}
\bibliography{jkNF12}

\end{document}